\documentclass[useAMS,usenatbib]{mn2e}
\usepackage{graphicx,amssym,color}
\newcommand{\Mpch}{\mbox{ $h^{-1}$ Mpc}}

\newcommand{\be}{\begin{equation}}
\newcommand{\ee}{\end{equation}}

\newcommand{\erf}{\textrm{erf}}

\def\ltsima{$\; \buildrel < \over \sim \;$}
\def\simlt{\lower.5ex\hbox{\ltsima}}
\def\gtsima{$\; \buildrel > \over \sim \;$}
\def\simgt{\lower.5ex\hbox{\gtsima}}

\renewcommand{\vec}[1]{ {\bmath #1} } 

\title{The Density and Pseudo-Phase-Space Density Profiles of CDM halos}

\author[Ludlow et al.] {\parbox{18cm}{
Aaron D. Ludlow$^{1,\star},$
Julio F. Navarro$^{2}$,
Michael Boylan-Kolchin$^{3,4}$,
Volker~Springel$^{5,6}$,
Adrian Jenkins$^{7}$,
Carlos S. Frenk$^{7}$,
and Simon D. M. White$^{3}$,
}\vspace{0.3cm}\\
$^{1}${Argelander-Institut f\"{u}r Astronomie, Auf dem H\"{u}gel 71,
D-53121 Bonn, Germany}\\
$^{2}${Dept. of Physics and Astronomy, University of
    Victoria, Victoria, BC, V8P 5C2, Canada}\\
$^3$Max-Planck-Institut f\"{u}r Astrophysik,
Karl-Schwarzschild-Stra\ss{}e 1, 85740 Garching bei M\"{u}nchen,
Germany\\
$^{4}${Center for Galaxy Evolution, 4129 Reines Hall, University of  
  California, Irvine, CA 92697, USA}\\
$^{5}${Heidelberg Institute for Theoretical Studies, 
Schloss-Wolfsbrunnenweg 35, 69118 Heidelberg, Germany}\\
$^{6}${Zentrum f\"{u}r Astronomie der Universit\"{a}t Heidelberg, ARI,
M\"{o}nchhofstr. 12-14, 69120 Heidelberg, Germany}\\
$^{7}${Institute for Computational Cosmology, Dept. of Physics, Univ. of
  Durham, South Road, Durham  DH1 3LE, UK}\\
}

\begin{document}

\maketitle 

\begin{abstract}
  Cosmological N-body simulations indicate that the
  spherically-averaged density profiles of cold dark matter halos are
  accurately described by Einasto profiles, where the logarithmic
  slope is a power-law of adjustable exponent, $\gamma\equiv d\ln
  \rho/d\ln r \propto r^\alpha$. The pseudo-phase-space density (PPSD)
  profiles of CDM halos also show remarkable regularity, and are well
  approximated by simple power laws, $Q(r)\equiv\rho/\sigma^3\propto
  r^{-\chi}$. We show that this is expected from dynamical equilibrium
  considerations, since Jeans' equations predict that the
  pseudo-phase-space density profiles of Einasto halos should resemble
  power laws over a wide range of radii. For the values of $\alpha$
  typical of CDM halos, the inner $Q$ profiles of equilibrium halos
  deviate significantly from a power law only very close to the
  center, and simulations of extremely high-resolution would be needed
  to detect such deviations unambiguously. We use an ensemble of halos
  drawn from the Millennium-II simulation to study which of these two
  alternatives describe best the mass profile of CDM halos.  Our
  analysis indicates that at the resolution of the best available
  simulations, both Einasto and power-law PPSD profiles (with
  adjustable exponents $\alpha$ and $\chi$, respectively) provide
  equally acceptable fits to the simulations. A full account of 
  the structure of CDM halos requires understanding how the shape parameters 
  that characterize departures from self-similarity, like $\alpha$ or $\chi$, are 
  determined by evolutionary history, environment or initial conditions.
\end{abstract}

\renewcommand{\thefootnote}{\fnsymbol{footnote}}
\footnotetext[1]{E-mail: aludlow@astro.uni-bonn.de} 

\section{Introduction}
\label{sec:intro}

The large dynamic range probed by current simulations of structure
formation allows for robust measurements of the internal structure of
large samples of dark matter halos spanning a wide range of
masses. The mass profile of CDM halos holds particular interest,
mainly because of its direct connection with key observational probes
of halo structure, such as disk galaxy rotation curves,
gravitational lensing measurements, and more recently because of the
possibility of observing the dark matter directly through its self-annihilation
signal, or in laboratory detectors.

\citet[][hereafter, NFW]{Navarro1996,Navarro1997} argued that
the simple 2-parameter formula
\begin{equation}
\rho(r)=\frac{\rho_c}{r/r_{s}(1+r/r_{s})^2},
\label{EqNFWProf}
\end{equation}
could be scaled to provide a good fit to the density profiles of
simulated halos. The two physical scaling parameters are $r_s$, the
radius where the logarithmic slope of the density profile,
$\gamma\equiv d\log \rho/d \log r$, equals $-2$ (the isothermal value)
and the characteristic density, $\rho_c$.  As discussed by NFW, the
parameters, $r_s$ and $\rho_c$, are not independent but rather
reflect the formation history of a given halo \citep[see also,][]
{Kravtsov1997,AvilaReese1999,Jing2000a,Bullock2001,Eke2001,Klypin2001}.

More recently, it has become clear that the mass profiles of CDM halos
are {\it not} strictly self-similar, as originally suggested by NFW,
but that a third parameter is actually required to describe their shape accurately
\cite[]{Navarro2004,Merritt2005,Merritt2006,Gao2008,Hayashi2008,Navarro2010}. These
studies have shown convincingly that profiles where the logarithmic slope is a simple
power-law of radius, $\gamma=-2\, (r/r_{-2})^{\alpha}$, provide excellent fits to 
simulated halo profiles when the shape parameter $\alpha$ is allowed to vary. This 
implies a density profile of the form
 \begin{equation}
\ln\biggr( \frac{\rho}{\rho_{-2}}\biggl)=-\frac{2}{\alpha}\biggr[\biggr(\frac{r}{r_{-2}}\biggl)^\alpha -1\biggl],
\label{EqEinastoProf}
\end{equation}
which we call the ``Einasto profile'' for short \citep[][]{Einasto1965}. The Einasto and 
NFW parameters are simply related by $r_{-2}=r_s$ and $\rho_{-2}=\rho_c/4$.

A second result that has received widespread attention has been the
fact that the pseudo-phase-space density (PPSD) profiles of
simulated dark matter halos follow simple power-laws with radius:
\begin{equation}
Q(r)\equiv \frac{\rho}{\sigma^3} = \frac{\rho_0}{\sigma_0^3}
\biggl(\frac{r}{r_0}\biggr)^{-\chi}.
\label{EqQ}
\end{equation}
This was originally reported by \citet{Taylor2001}, and has been
confirmed by a number of subsequent studies, albeit with some debate
over the actual value of the best-fit power-law exponent \citep[see,
e.g.,][]{Rasia2004,Dehnen2005,Faltenbacher2007,Vass2009,Ludlow2010}. Although
originally shown to hold for the total velocity dispersion profile,
eq.~(\ref{EqQ}) also holds when $\sigma$ is replaced with
$\sigma_r$, the radial velocity dispersion
\citep[e.g.,][]{Dehnen2005,Navarro2010}. (We hereafter use $Q$ when
referring to the total pseudo-phase-space density profile, and
$Q_r$ for its radial analog, $\rho/\sigma_r^3$.)

As discussed by \citet{Taylor2001}, the power-law nature of $Q$ does
not fully determine the halo density profile. Indeed, a wide range of
different density profiles are consistent with this constraint, even
for spherically symmetric, isotropic systems in dynamical
equilibrium. Well-behaved solutions, however, have only two possible
asymptotic inner behaviours, one where the central density diverges
like $\rho \propto r^{2\chi-6}$ (the singular isothermal sphere is an
example, for $\chi=2$), and another ``critical'' solution where the
cusp asymptotic slope is much shallower; $\gamma\rightarrow -2\chi/5$
as $r\rightarrow 0$. \citet{Taylor2001} showed that the ``critical''
solution for $\chi=1.875$ (the value predicted by the self-similar
secondary infall solution studied by \citealt{Bertschinger1985})
closely resembles the NFW profile over the radial range resolved by
their simulations. 

The inner asymptotic behaviour predicted by eqs.~(\ref{EqEinastoProf})
and (\ref{EqQ}) for the density profile is therefore different (Einasto
profiles have a finite-density core rather than a cusp), implying that
simulations should be able to discriminate between the two. Indeed, the $Q(r)$
profiles of Einasto halos are {\it not} in general power-laws
\citep[][]{Ma2009}, which implies that halos cannot satisfy both
eqs.~(\ref{EqEinastoProf}) and ~(\ref{EqQ}) at once and that departures
from either Einasto fits or power-law $Q$ profiles should become
apparent in simulations of adequate resolution.

Deviations from power laws in the {\it outer} $Q$ profiles of
simulated halos have already been reported by \citet{Ludlow2010}, who
argue that such departures are actually expected in regions near the
virial boundary separating the inner equilibrium region from the
unrelaxed infalling envelope of the halo. Whether similar deviations are also
present in the inner regions of halos has not yet been studied in
detail.
 
We address these issues here using a sample of equilibrium CDM halos
selected from the Millennium-II (MS-II) simulation
\citep{Boylan-Kolchin2009}. We begin in Section \ref{sec:sims} with a brief
description of our simulations and analysis techniques. Our main
results are presented in Section \ref{SecRes}; density profiles are
presented in Section \ref{SecRhoProf} and $Q_r$ profiles in
Section \ref{SecQProf}.  Since both Einasto and power-law PPSD profiles are
incomplete dynamical models unless the velocity anisotropy profile is
specified we analyze this in Section \ref{SecBetaProf}. In
Section \ref{SecRhoPPSD} and \ref{SecEinPPSD} we compare the Einasto and
power-law PPSD models and directly asses their ability to accurately
describe the mass profiles of simulated halos. We end with a brief
summary of our main conclusions in Section \ref{SecConc}.

\section{Numerical Simulations}
\label{sec:sims}

\subsection{Cosmological model}
\label{ssec:cosmology}

Our analysis uses group and cluster halos identified in the MS-II, a $\sim
10^{10}$-particle cosmological simulation of the evolution of dark
matter in a 100\Mpch{} box. The run adopted a standard
$\Lambda$CDM cosmogony with the same parameters as the Millennium
simulation presented by \citet{Springel2005a}: $\Omega_{\rm M}=0.25$,
$\Omega_{\Lambda}=1-\Omega_{\rm M}=0.75$, $n_s=1$, $\sigma_8=0.9$, and
a Hubble constant $H_0\equiv H(z=0)=100\, h$ km s$^{-1}$ Mpc$^{-1}=73$
km s$^{-1}$ Mpc$^{-1}$, but has better mass and spatial
resolution. Interested readers may find further details in
\citet{Boylan-Kolchin2009}.

\subsection{Halo selection}
\label{ssec:selection}

We base our results on a sample of well-resolved equilibrium halos in
MS-II. We focus on halos where ${\rm N_{200}}$, the number of
particles within the virial radius\footnote{The virial radius,
  $r_{200}$, defines the mass of a halo, $M_{200}$, as that of a
  sphere, centered at the potential minimum, whose enclosed density is
  $200$ times the critical density for closure, $\rho_{\rm
    crit}=3H^2/8\pi G$.  $V_{200}=(GM_{200}/r_{200})^{1/2}$ is the
  halo's virial velocity. }, exceeds $5\times 10^5$. This corresponds
to group and cluster halos with virial masses above $\sim 3.44\times
10^{12} h^{-1} {\rm M}_{\odot}$.

In addition we require our halos to satisfy the set of relaxation
criteria introduced by \citet{Neto2007} in order to minimize the
impact of transient departures from equilibrium on halo
profiles. These criteria include limits on (i) the fraction of a halo's
virial mass in self-bound substructures, $f_{\rm sub}=M_{\rm
  sub}(r<r_{200})/M_{200}<0.07$; ii) the offset between a halo's
center of mass and its true center (defined by the particle with the
minimum potential energy), $d_{\rm off} = |\vec{r}_{\rm
  CM}-\vec{r}_{\rm cen}|/r_{200}<0.05$, and iii) the virial ratio of
kinetic to potential energies, $2K/|\Phi|<1.3$. All dynamical
quantities have been computed in the halo rest frame. With these
selection criteria our sample consists of 440 halos, more than half of
the total number of halos (790) in the same mass range.

%%%%%%%%%%%%%%%%%%%%%%%%%%%%%%%%%%%%%%%%%%%%%%%%%%%%%%%%%%%%%%                                
\begin{figure*} 
\begin{center} \resizebox{18cm}{!}{\includegraphics{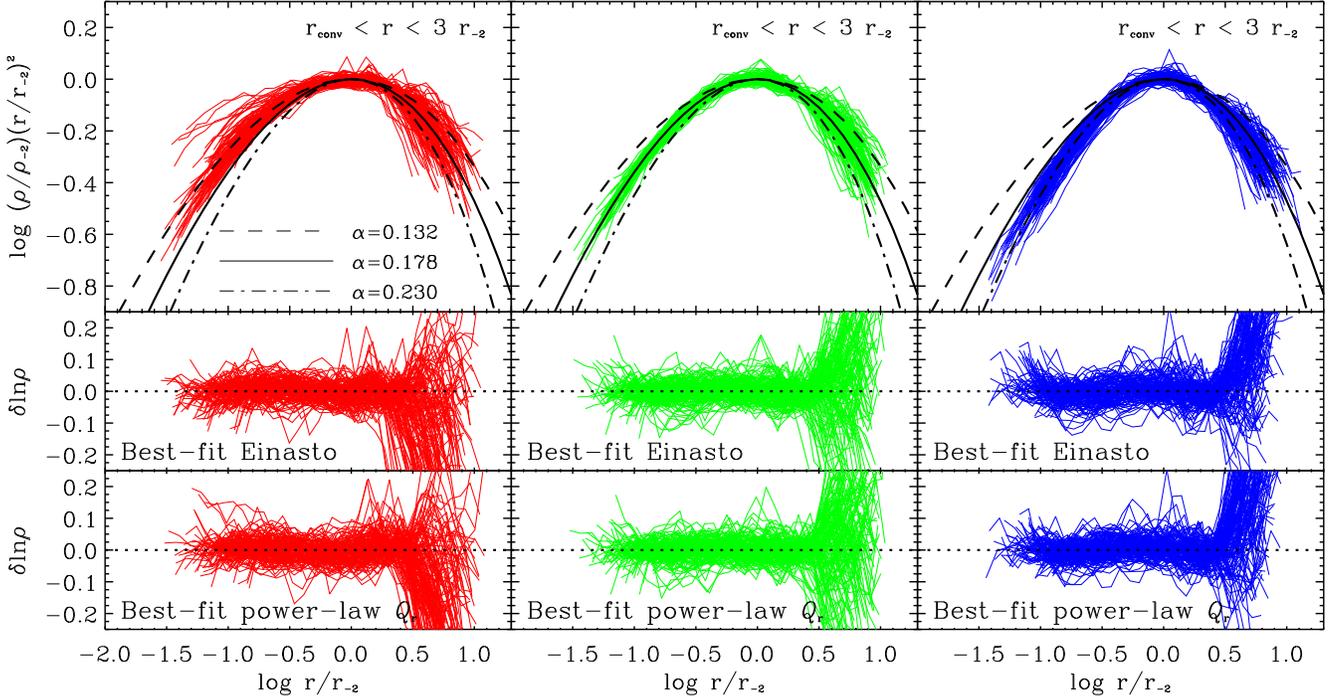}}
\end{center} 
\caption{Spherically-averaged density profiles of all halos in our
  sample. All profiles are plotted over the radial range $r_{\rm
    conv}< r< r_{200}$. Radii have been scaled by $r_{-2}$; densities
  by $\rho_{-2}\equiv \rho(r_{-2})$. Density estimates have been
  multiplied by $r^2$ in order to increase the dynamic range of the
  plot so as to highlight differences between halos. The full sample
  of halos has been divided into three subsamples according to the
  best-fit value of the Einasto parameter $\alpha$: red curves (left
  panels) correspond to halos with $\alpha\leq 0.161$; green curves
  (middle panels) to halos with $0.161<\alpha \leq 0.195$ and blue
  curves (right panels) to halos with $\alpha > 0.195$. There
  are equal numbers of halos in each subsample. To illustrate the role
  of the shape parameter, $\alpha$, Einasto profiles with
  $\alpha=0.132$ (dashed), $0.178$ (solid) and $0.230$ (dot-dashed)
  are shown in the top panels. The middle
  panels are residuals from the best-fit Einasto profile
  with adjustable $\alpha$; bottom panels are the residuals from the
  best-fit power-law $Q_r$ ``critical'' model (see
  Sec.~\ref{SecRhoPPSD}). In all cases fits are carried out over the
  radial range $r_{\rm conv} < r < 3 \ r_{-2}$.}
\label{FigPLRhoProf} 
\end{figure*}
%%%%%%%%%%%%%%%%%%%%%%%%%%%%%%%%%%%%%%%%%%%%%%%%%%%%%%%%%%%%%%          

\subsection{Analysis}
\label{ssec:analysis}

Our analysis deals with the spherically-averaged density, $\rho$, and
velocity dispersion, $\sigma$, profiles of each of our 440 dark matter
halos. Each profile is built out of $25$ spherical shells equally
spaced in $\log_{10} r$ spanning the range $r_{\rm conv}\leq r\leq
r_{200}$. Here $r_{\rm conv}$ is the ``convergence radius'' defined by
\citet[]{Power2003}, where circular velocities converge to better than
$\sim$10\%. Each spherical shell is centered at the particle with the
minimum potential energy, which we identify with the halo center.

For each radial bin we estimate the dark matter mass-density by
dividing the total mass of the shell by its volume; the velocity
dispersion is defined such that $\sigma^2$ is two times the specific
kinetic energy in the shell, measured in the halo rest
frame. Analogously, we compute the radial velocity dispersion,
$\sigma_r^2$, in terms of the kinetic energy in radial motions within
each shell. The velocity anisotropy parameter is defined by
$\beta=1-\sigma_{\rm tan}^2/2\sigma_{r}^2$, where $\sigma_{\rm
  tan}^2=\sigma^2-\sigma_r^2$.  The $\rho$ and $\sigma_r$ values in
each radial shell are also used to estimate the pseudo-phase-space
density profile, $Q_r(r)\equiv\rho/\sigma_r^3$. We choose to focus our
analysis on the {\em radial} $Q_r$ profiles (rather than the total
$Q\equiv\rho/\sigma^3$) as this simplifies the analysis of Jeans'
equations, in which the $\sigma_r$ and $\beta(r)$ terms separate.

%%%%%%%%%%%%%%%%%%%%%%%%%%%%%%%%%%%%%%%%%%%%%%%%%%%%%%%%%%%%%%                                
\begin{figure*} 
\begin{center} \resizebox{18cm}{!}{\includegraphics{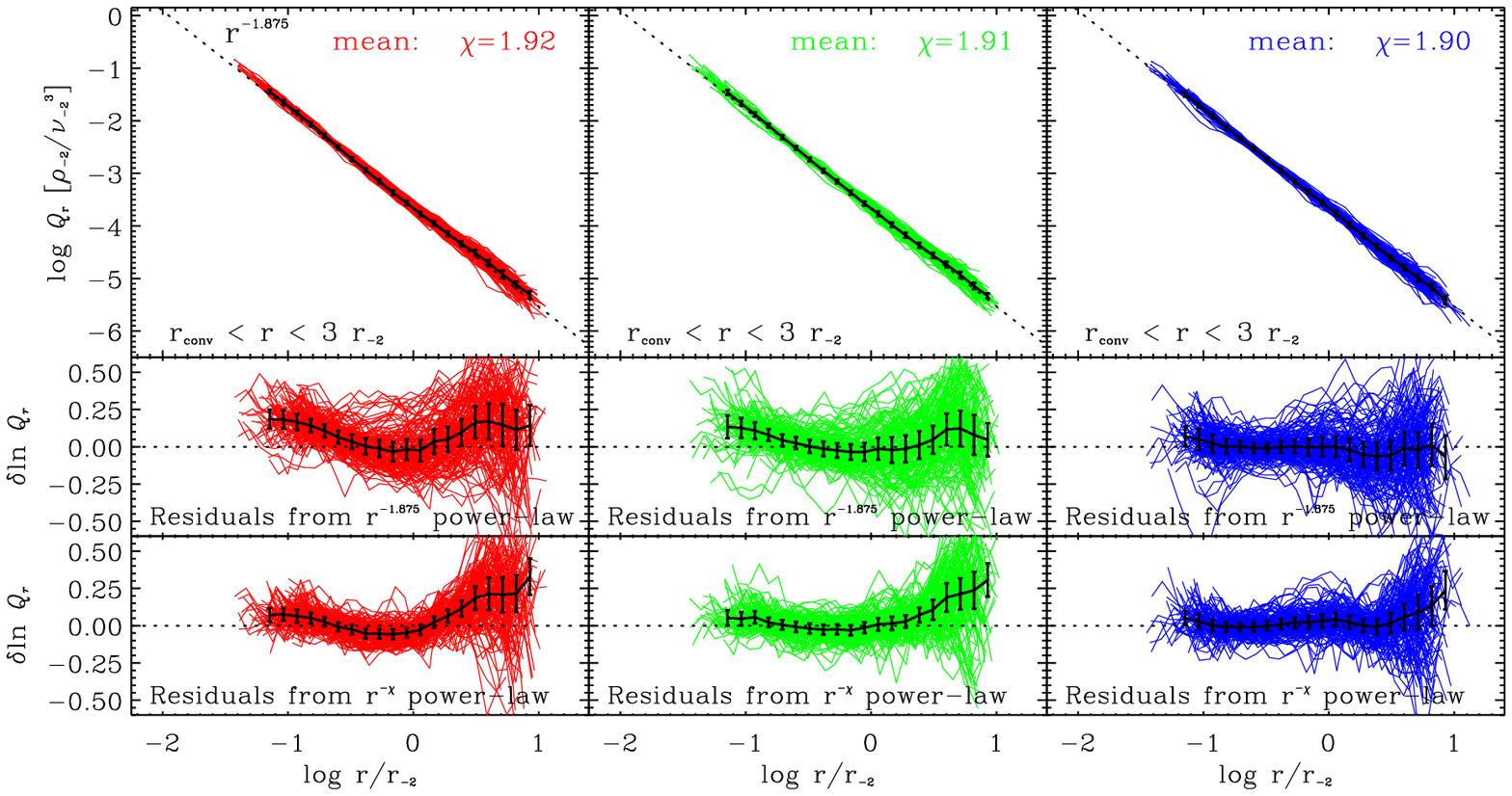}} 
\end{center} 
\caption{Radial pseudo-phase-space density profiles,
  $Q_r=\rho/\sigma_r^3$, for all halos in our sample.  As in
  Fig.~\ref{FigPLRhoProf}, halos are subdivided into three separate
  subsamples according to the value of their best-fit Einasto
  parameter $\alpha$. Radii have been scaled by $r_{-2}$, densities by
  $\rho_{-2}$ and velocities by $\nu_{-2}=\sqrt{G\rho_{-2}\,
    r_{-2}^2}$. The mean profiles and one-sigma scatter are shown as
  solid lines with error bars.  The dotted curves in the top panels
  show a Bertschinger $r^{-1.875}$ power-law to guide the eye.  The
  middle panels plot the residuals from the Bertschinger
  law normalized to the mean value of $\rho_{-2}/\nu_{-2}^3$
  for all halos in each sample. Lower panels show the residuals from the best-fit power-law,
  $r^{-\chi}$, with free-floating exponent, $\chi$.  Fits are limited
  to the radial range $r_{\rm conv} < r < 3\, r_{-2}$. The
  $\alpha$-dependence of the $Q_r$ profiles is clearly evident from
  the systematic differences in the shape of the residual curves shown in
  the middle panels.}
\label{FigQProf} 
\end{figure*}
%%%%%%%%%%%%%%%%%%%%%%%%%%%%%%%%%%%%%%%%%%%%%%%%%%%%%%%%%%%%%%          

\section{Results}
\label{SecRes}
\subsection{Density profiles}
\label{SecRhoProf}

Figure~\ref{FigPLRhoProf} shows the spherically-averaged density
profiles for halos in our sample, with the residuals from best-fit
Einasto profiles shown in the middle panels.  All profiles have been
plotted from $r_{200}$ down to the innermost resolved radius, $r_{\rm
  conv}$.  Densities have been multiplied by $r^2$ in order to enhance
the dynamic range of the graph and to highlight halo-to-halo
differences. Einasto fits are limited to the radial range $r_{\rm
  conv} < r < 3 \ r_{-2}$, so that their parameters are not unduly
influenced by regions where crossing times are long, and dynamical equilibrium 
may not hold \citep{Ludlow2010}. Radii have been
scaled to $r_{-2}$, the radius at which the logarithmic slope is equal
to the isothermal value, $-2$, and densities by $\rho_{-2}
\equiv\rho(r_{-2})$.

We assess the quality of fits to the density profiles using the
following figure-of-merit function,
\begin{equation}
\psi^2 = \frac{1}{N_{\rm bin}} \sum_{i=1}^{\rm Nbin} (\ln \rho_i - \ln \rho_i^{\rm model})^2.
\label{eq:FoM}
\end{equation}
All halos shown in Fig.~\ref{FigPLRhoProf} have $\psi_{\rm min}<0.1$.

In order to highlight the differences in profile shape we have split
our halo sample into three subsamples, each containing an equal number
of halos: halos with $\alpha\leq 0.161$ are shown in red; green curves
show those with $0.161 < \alpha \leq 0.195$, and blue curves are used
for the rest. We adopt the same colour coding and halo samples in
subsequent plots. For illustration, we also show three Einasto
profiles with different values of the shape parameter, corresponding
to the mean value of each halo subsample: $\alpha=0.132$ (dashed
curve), $\alpha=0.178$ (solid black curve), and $\alpha=0.230$ (dot-dashed
curve).

The small residuals from best Einasto fits (typically less than 10\%
throughout the fitted region; see middle panels of
Fig.~\ref{FigPLRhoProf}) confirm that eq.~(\ref{EqEinastoProf}) provides
an excellent description of the density profiles of CDM halos down to
the innermost resolved radius.  Further, Figure~\ref{FigPLRhoProf}
also demonstrates that there is genuine variation in profile shape
from halo to halo: profiles of different halos ``curve'' differently,
yielding best-fit $\alpha$ parameter values that range between roughly
$0.1$ and $0.25$. The mass profiles of CDM halos are therefore {\it
  not} strictly self-similar, and $\alpha$ is a structural ``shape''
parameter genuinely needed to describe accurately the mass profile of
CDM halos \citep[see][for a full discussion]{Navarro2010}.

\subsection{Pseudo-phase-space density profiles}
\label{SecQProf}

Fig.~\ref{FigQProf} shows the (radial) pseudo-phase-space density
profiles, $Q_r$, of the halos in our sample, grouped as in
Fig.~\ref{FigPLRhoProf}. In order to take out the halo mass dependence,
$Q_r$ profiles have been scaled radially by $r_{-2}$, and
vertically by $\rho_{-2}/\nu_{-2}^3$, where $\nu_{-2}=\sqrt{G \,
  \rho_{-2} \, r_{-2}^2}$ is the characteristic velocity implied by
the scale parameters $\rho_{-2}$ and $r_{-2}$.  Our results confirm
the striking power-law nature of $Q_r$ profiles previously reported in
the literature. For reference, we also show the power law, $Q_r\propto
r^{-1.875}$ (shown by a black dotted line), predicted by the
self-similar solution of \citet{Bertschinger1985}.

%%%%%%%%%%%%%%%%%%%%%%%%%%%%%%%%%%%%%%%%%%%%%%%%%%%%%%%%%%%%%%                                
\begin{figure*} 
\begin{center} \resizebox{18cm}{!}{\includegraphics{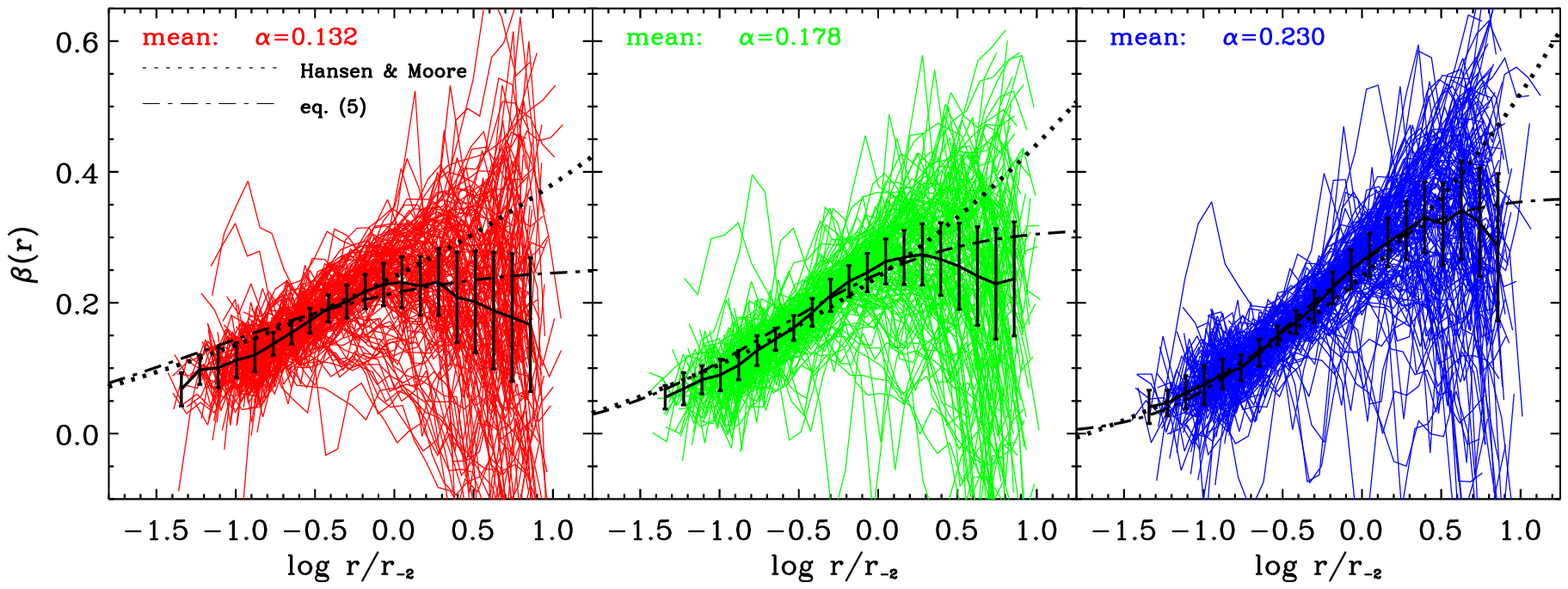}}
\end{center} 
\caption{Velocity anisotropy profiles, $\beta=1-\sigma_{\rm tan}^2/2\sigma_r^2$, 
  for all halos in our sample. Profiles have been grouped according to 
  the best-fit Einasto parameter, as in Figure~\ref{FigPLRhoProf}. Solid black 
  lines with error bars show the median anisotropy profile and one-sigma
  dispersion for each halo subsample. The dot-dashed curves shown in each panel 
  show the radial dependence of $\beta(r)$ expected from 
  eq.~(\ref{eq:betagamma}) assuming the average Einasto density profile; dotted curves show 
  $\beta(r)$ expected from \citet{Hansen2006} for the same $\rho(r)$.}
\label{FigBetaProf} 
\end{figure*}
%%%%%%%%%%%%%%%%%%%%%%%%%%%%%%%%%%%%%%%%%%%%%%%%%%%%%%%%%%%%%%     

In order to emphasize the differences between halos, we show in the
middle panels of Fig.~\ref{FigQProf} the residuals from the
Bertschinger power law normalized to the mean value of
$\rho_{-2}/\nu_{-2}^3$ for all halos in each sample.  Although the
residuals are in general small, the ``curving'' shape of their radial
dependence provides a strong indication that the $Q_r$ profiles of
simulated CDM halos deviate systematically from a simple power-law of
fixed exponent $\chi$.

As discussed by \citet{Ludlow2010}, the {\it outer} ($r\gg r_{-2}$)
upturn in the residuals is most likely associated with the transition
from the inner, relaxed parts, to the unrelaxed outer parts, where
infalling material has not yet had time to phase-mix with the main
body of the halo. Such an upturn is present also in the self-similar
solution of \citet{Bertschinger1985}, and may be a
general feature of the outer $Q_r$ profiles of CDM halos. Because of
this, we have chosen the radial range $r_{\rm conv} <r<3\, r_{-2}$ for
all the fits we report here.

Interestingly, the {\it inner} $Q_r$ profiles ($r \simlt 0.1 \,
r_{-2}$) also deviate from the $r^{-1.875}$ power law in a way that
clearly depends on $\alpha$. Note, for example, that at the innermost
point the residuals of the lowest-$\alpha$ halos (leftmost middle panel
in Fig.~\ref{FigQProf}) are substantially larger than those of the
largest-$\alpha$ halos (rightmost middle panel).  We have explicitly
verified that this $\alpha$-dependence is {\it not} caused by grouping
halos of different mass, nor by differences in numerical
resolution. It is also unlikely to be due to anisotropies in the
velocity distribution, since, as we show below, all halos in our
sample are nearly isotropic at radii this close to the center. We have
also checked that this $\alpha$ dependence is not specific to our
choice of $Q_r$; the total pseudo-phase-space density profiles,
$Q=\rho/\sigma^3$, follow similar trends with $\alpha$.

These results imply that no single power-law can reproduce the radial
dependence of the pseudo-phase-space density; if $Q_r$ does indeed
follow a power-law of radius, then the exponent $\chi$ must vary from
halo to halo.  We show this explicitly in the bottom panels of
Figure~\ref{FigQProf}, where we plot the residuals from the best-fit
power-law when the exponent $\chi$ is allowed to vary. The small
residuals over the fitted radial range $r_{\rm conv} < r < 3\,
r_{-2}$) indicate that a power-law with adjustable $\chi$ provides a
remarkably accurate description of the inner $Q_r$ profiles of CDM
halos.

Do Einasto profiles provide a better description of the spherically
averaged structure of CDM halos than power-law $Q_r$ profiles, or
vice versa? Because of different dimensionality, we cannot compare
directly the goodness of fits to the $\rho$ and $Q_r$ profiles shown
in Figs.~\ref{FigPLRhoProf} and \ref{FigQProf}. One way to make
progress is to compute the PPSD profiles corresponding to Einasto
halos, or, alternatively, to compute the density profiles of power-law
PPSD models and compare them with the simulations. This may be
accomplished by assuming dynamical equilibrium and solving Jeans'
equations to link the $\rho$ and $Q_r$ profiles, a procedure that,
however, requires an assumption regarding the radial dependence of the
velocity anisotropy, $\beta(r)$. We turn our attention to that issue
next.

\subsection{Velocity anisotropy}
\label{SecBetaProf}

Velocity anisotropy profiles for all halos in our sample are shown in
Figure~\ref{FigBetaProf}, after rescaling all radii to $r_{-2}$. Halos
have been grouped according to the value of the best-fit $\alpha$
parameter, as in Figure~\ref{FigPLRhoProf}. Solid lines with error
bars show the median $\beta(r)$ profile of each group and the associated one-sigma
scatter. The velocity anisotropy profiles exhibit a characteristic
shape: they are isotropic near the center, become increasingly radial
with increasing distance, before leveling off or becoming less
anisotropic in the outskirts.

\begin{figure} 
\begin{center} \resizebox{8cm}{!}{\includegraphics{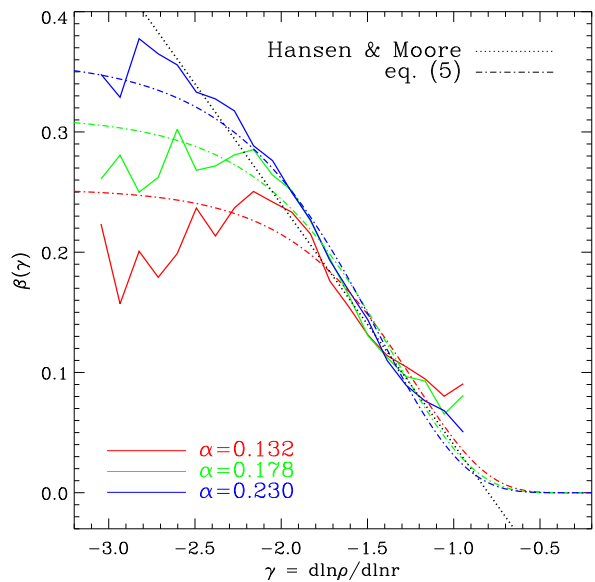}}
\end{center} 
\caption{Median density slope-velocity anisotropy relation for the
  three groups of halos shown in previous figures. The linear $\beta -
  \gamma$ relation proposed by \citet{Hansen2006} is shown as a dotted
  line; a relation of the form given in eq.~(\ref{eq:betagamma}) is
  shown separately for each sample using dot-dashed curves. 
  There is some evidence that the mean
  $\beta-\gamma$ relation departs from the \citet{Hansen2006} result
  in the outer regions of our halos. See text for
  further discussion.}
\label{FigBetaGamma} 
\end{figure}

Figure~\ref{FigBetaGamma} shows the logarithmic slope-velocity
anisotropy ($\gamma$ vs $\beta$) relation for the median profiles of
each halo subsample using the same colour coding as previous plots. The
dotted line shows the linear $\beta(\gamma)$ relation proposed by
\citet{Hansen2006}, which reproduces our simulation data well within
the halo scale radius $r_{-2}$ (i.e., for $\gamma>-2$).

The data in Fig.~\ref{FigBetaGamma} also suggests that the
$\beta(\gamma)$ relation deviates from the Hansen \& Moore fit in
the outer regions (where $\beta$ tends to a constant rather than the
steadily increasing radial anisotropy predicted by Hansen \&
Moore). This dependence is captured well by the function
\begin{equation}
\beta(\gamma) = \frac{\beta_{\infty}}{2} \left(1 + \erf(\ln[(A\gamma)^2])\right),
\label{eq:betagamma}
\end{equation}
as shown by the dot-dashed curves in Figure~\ref{FigBetaGamma}. The
values of $\beta_{\infty}$ and $A$ depend weakly but systematically on
$\alpha$, and are listed in Table~\ref{tab1}. The radial $\beta$
profiles corresponding to eq.~(\ref{eq:betagamma}) are also shown with
dot-dashed curves in each panel of Figure~\ref{FigBetaProf}, assuming
that $\gamma(r)$ is given by the average Einasto profile of each
grouping.

\subsection{Density profiles from  power-law PPSD profiles}
\label{SecRhoPPSD}

Once the radial dependence of the velocity anisotropy has been
specified, equilibrium density profiles consistent with a power-law
$Q_r(r)$ profile may be obtained from Jeans' equation. Following
\citet{Dehnen2005}, we write Jeans' equation as
\begin{eqnarray}
  \left(\gamma^\prime -\frac{6}{5}\,\beta^\prime\right) &+&
  \frac{2}{3}\left(\gamma+\chi+\frac{3}{2}\right)
  \left(
  \gamma+\frac{2}{5}\chi+\frac{6}{5}\,\beta
  \right) \nonumber \\
 &=& -\frac{3}{5}\,\kappa_{-2}\, x^{2-2\chi/3} y^{1/3},
  \label{EqJeans1}
\end{eqnarray}
where $y\equiv\rho/\rho_{-2}$; $x\equiv r/r_{-2}$;
$\kappa_{-2}\equiv 4\pi G\rho_{-2} r_{-2}^2/\sigma_{r,-2}^2$ is a
measure of the velocity dispersion in units of the ``natural''
velocity scale of the halo at $r_{-2}$, and we have assumed that
$Q_r(r)$ is a power law of exponent $\chi$. The logarithmic slope,
$d\ln\rho/d\ln r$ is denoted by $\gamma$, and primes indicate
derivatives with respect to $\ln x$. Once $\beta(r)$ is specified, the
solution set of eq.~(\ref{EqJeans1}) for given $\chi$ is fully
characterized by $\kappa_{-2}$.

%%%%%%%%%%%%%%%%%%%%%%%%%%%%%%%%%%%%%%%%%%%%%%%%%%%%%%%%%%%%%%                                
%%%%%%%%%%%%%%%%%%%%%%%%%%%%%%%%%%%%%%%%%%%%%%%%%%%%%%%%%%%%%%          

%%%%%%%%%%%%%%%%%%%%%%%%%%%%%%%%%%%%%%%%%%%%%%%%%%%%%%%%%%%%%%                                
\begin{figure*} 
\begin{center} \resizebox{18cm}{!}{\includegraphics{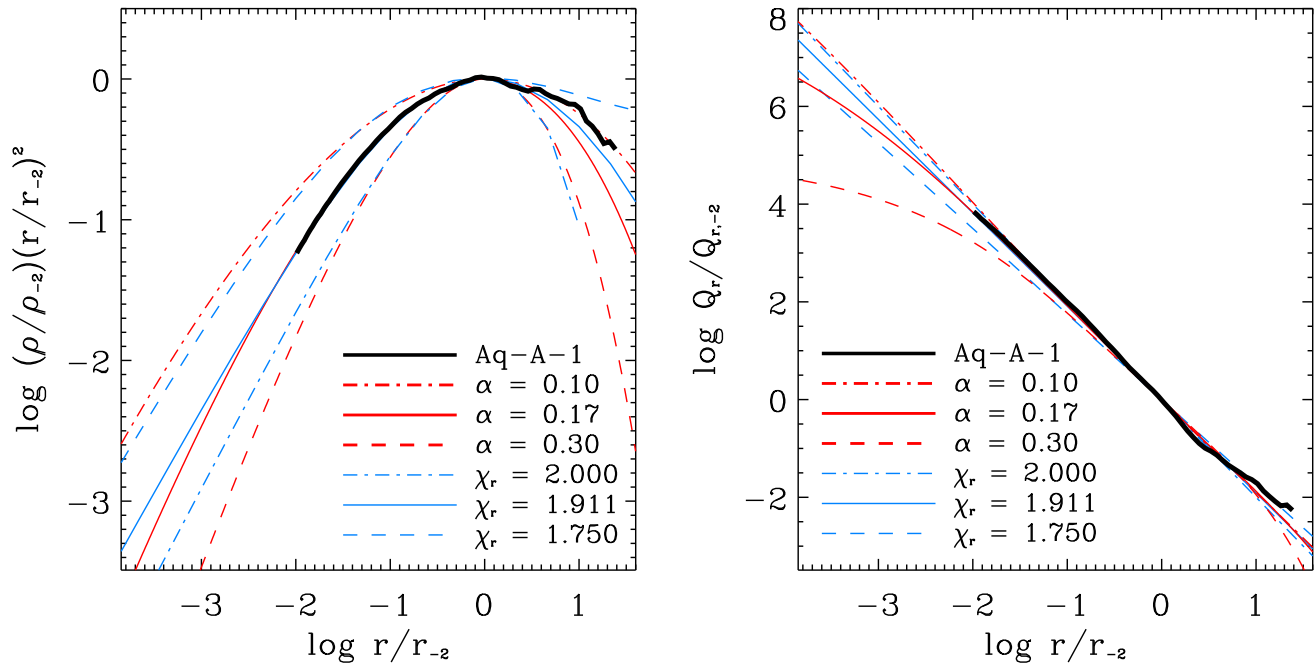}}
\end{center} 
\caption{{\it Left:} Einasto density profiles compared with
  ``critical'' profiles computed from Jeans' equation assuming power-law PPSD models. Three
  different values of $\alpha$ and of $\chi$ are shown, matched to
  highlight the similarity in the density profiles over a large radial
  range. {\it Right:} Radial PPSD profiles of Einasto halos compared
  with power-law models. As in the left panel, the values of $\alpha$
  and $\chi$ of the three curves have been chosen so as to highlight
  the similarity of each pair of curves over a wide radial range. In
  both panels the thick solid curve in black shows the profile
  corresponding to the billion-particle Aq-A-1 halo, the
  highest-resolution CDM halo published to date. At the resolution of
  this halo, or lower, both Einasto and power-law PPSD models provide
  equally acceptable descriptions of the spherically-averaged
  structure of CDM halos.}
\label{FigJeansSoln} 
\end{figure*}
%%%%%%%%%%%%%%%%%%%%%%%%%%%%%%%%%%%%%%%%%%%%%%%%%%%%%%%%%%%%%%          

It is instructive to consider the special case $\beta(r)=0$, for which
three central asymptotic behaviours are possible: (i) a steep central
cusp, $\gamma\rightarrow 2\chi-6$; (ii) a central ``hole'' where
$y(0)=0$; or (iii) a ``critical'' solution with a shallow central
cusp, $\gamma\rightarrow -2\chi/5$.  The latter may be thought of as
the limiting case where the radius of the central ``hole'' solution
goes to zero, and corresponds to a maximally-mixed state for given
halo binding energy and mass \citep{Taylor2001}.

Given that steep central cusps such as those in (i) are firmly ruled
out by the data (see, e.g., \citealt{Navarro2010}), the ``critical''
solutions (iii) are the only viable density profiles consistent with
the power-law PPSDs constraint. Although these considerations are
strictly true only for isotropic systems, similar conclusions apply
once $\beta(r)$ is specified. This is especially true considering that
halos are very nearly isotropic close to the center (see
Sec.~\ref{SecBetaProf}). We shall hereafter adopt the ``critical''
solutions (evaluated numerically) as the density profiles
corresponding to a power-law PPSD profile for a given value of $\chi$.

The left panel of Figure~\ref{FigJeansSoln} shows (in blue) the ``critical''
density profiles for three different values of $\chi$, and compares
them to Einasto profiles. The values of $\alpha$ of the three Einasto
profiles shown have been chosen to match as closely as possible the
profiles corresponding to the PPSD models. 

Clearly, for every value of $\chi$ it is possible to choose a value of
$\alpha$ that matches the resulting density structure over a very wide
range in radius. Deviations are only seen in the very inner regions,
at less than 0.1\% of the scale radius and therefore well outside the
convergence region of any published halo simulation to date. For
example, the black solid line in Fig.~\ref{FigJeansSoln} shows the
density profile of the billion-particle Aq-A-1 halo, whose convergence
radius is roughly $0.01\, r_{-2}$ \citep{Navarro2010}. An Einasto
profile with $\alpha=0.17$ and a ``critical'' solution for
$\chi=1.911$ match the profile of this halo indistinguishably well.

\subsection{PPSD profiles from Einasto profiles}
\label{SecEinPPSD}

Pseudo-phase-space density profiles corresponding to Einasto models
can also be obtained by solving Jeans' equation, which may be written
as follows:
\begin{equation}
\frac{d\ln\zeta_r^2}{d\ln r}-\zeta_r^2=\gamma+\frac{d\ln V_c^2}{d\ln r}+2\beta,
\label{EqJeans2}
\end{equation}
where $\zeta_r^2\equiv V_c^2/\sigma_r^2$. With $\beta(r)$ set by
eq.~(\ref{eq:betagamma}), the right hand side of this equation is fully
specified by $r_{-2}$, $\rho_{-2}$, and $\alpha$. Solutions may
therefore be found after choosing the value of $\zeta_r$ (or its
logarithmic derivative) at some fiducial radius, such as
$\zeta_{-2}=\zeta_r(r_{-2})$.   The shape of the $Q_r$ profile is dictated by $\zeta_{-2}$, and it
is not generally a power law \citep{Ma2009}.  

Again, insight may be gained by considering the limiting behaviour of
the isotropic solutions. Because Einasto profiles have finite central
densities, $\sigma_r$ must approach a constant at the center.  The
central velocity dispersion may be found numerically, and turns out to
be quite insensitive to $\zeta_{-2}$ (for given $\rho_{-2}$, $r_{-2}$,
and $\alpha$), provided that the $\zeta_{-2}$ is greater than about a
tenth. (All of our halos are comfortably in that regime; indeed, the
median value of $\zeta_{-2}$ for all of our halos is $1.274$.) The finite
value of the central velocity dispersion and its near invariance with
$\zeta_{-2}$ also imply that the shape of the inner velocity
dispersion profile is quite insensitive to the actual value of
$\zeta_{-2}$. 

Therefore there is, in practice, a unique $Q_r$ profile that corresponds
to an Einasto profile of given $\alpha$, which we identify with the
single solution that is well behaved at all radii. This may be found
by setting $d\ln\zeta_r^2/d\ln r = \alpha$ at $r=\infty$; for
$\alpha=0.175$, for example, this implies $\zeta_{-2}=1.265$.

The right panel of Figure~\ref{FigJeansSoln} shows (in red) the PPSD
profiles of Einasto halos, for three different values of $\alpha$. For
$\alpha=0.1$ and $0.17$ the corresponding PPSD profiles are very well
approximated by power laws over the whole plotted radial range. Only
for larger values of $\alpha$, such as $0.3$, are clear deviations
from a power law noticeable. Even in this case, however, these are
only evident in regions well inside $1\%$ of the scale radius
$r_{-2}$, and therefore outside the converged region of the
highest-resolution simulation of a CDM halo published to date, the
Aq-A-1 halo (shown by a solid black line). Deciding whether power-law
PPSD models match CDM halos better than Einasto profiles, or
vice versa, seems to require simulations of even better resolution than
Aq-A-1.

\subsection{Power-law PPSD vs Einasto density profile fits}

The results above suggest that both Einasto profiles and power-law
$Q_r$ models provide equally good representations of the
spherically averaged structure of simulated CDM halos, in spite of
making very different predictions for their {\it asymptotic} inner structure. 

This conclusion may be verified quantitatively by fitting the
``critical'' solutions introduced in Sec.~\ref{SecRhoPPSD} to the
density profiles of all halos in our sample and comparing them with
Einasto fits. Residuals from the best fits obtained after varying
$\chi$ are shown in the bottom panels of Figure~\ref{FigPLRhoProf},
and are quite clearly indistinguishable from those obtained by fitting
Einasto laws with adjustable $\alpha$ (shown in the middle panels of
the same figure).

\begin{figure} 
\begin{center} \resizebox{8cm}{!}{\includegraphics{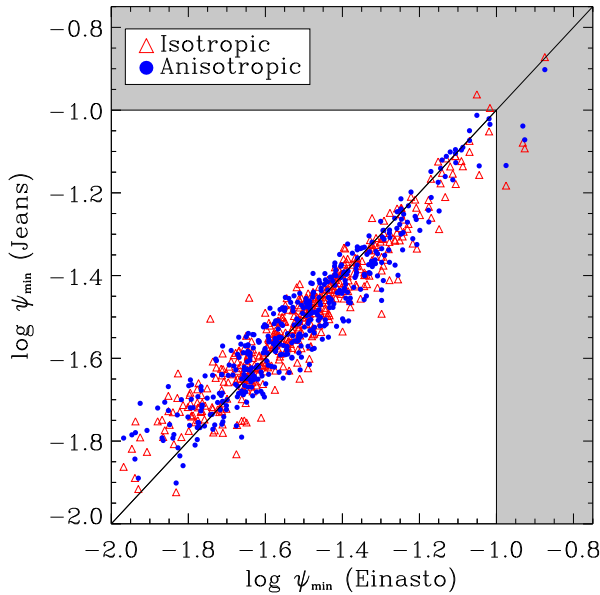}}
\end{center} 
\caption{Figure of merit, $\psi_{\rm min}$, of the best density profile fits to
  halos in our sample.  Values on the y-axis are determined from the
  best-fitting ``critical'' power-law $Q_r$ model; those in the
  abscissa correspond to the best-fitting Einasto profiles. In all
  cases, fits are limited to the radial range $r_{\rm conv}<r<3\
  r_{-2}$ and have the same number of degrees of freedom: two scale
  parameters, $\rho_{-2}$ and $r_{-2}$, and one shape parameter,
  $\chi$ or $\alpha$. Halos in the shaded grey area have been omitted
  from the analysis.}
\label{FigFoM} 
\end{figure}

Further quantitative evidence is provided in Fig.~\ref{FigFoM}, where
we plot the figure of merit, $\psi_{\rm min}$, of the best Einasto
fits compared with that obtained from the best-fitting critical
solution. Open and filled symbols in this figure are used to denote
cases where, respectively, either isotropic or anisotropic (i.e.,
$\beta$ given by eq.~(\ref{eq:betagamma})) critical solutions have been
used in the fitting procedure. The close resemblance of the results
obtained with either assumption implies that our conclusions are
largely insensitive to our assumption about the radial dependence of
the velocity anisotropy.

\begin{figure} 
\begin{center} \resizebox{8cm}{!}{\includegraphics{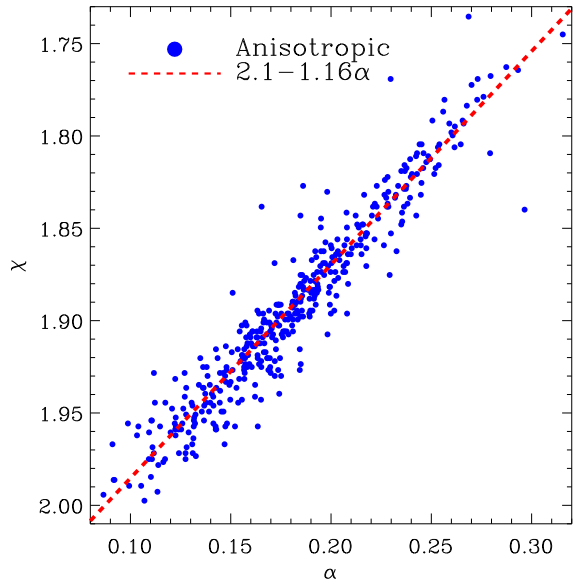}}
\end{center} 
\caption{The relation between best-fit shape parameters, $\chi$ and
  $\alpha$, derived from fits to the spherically averaged density
  profiles using either Einasto profiles or ``critical'' profiles
  corresponding to power-law PPSD models. All fits are carried out
  over the radial range $r_{\rm conv} < r < 3\, r_{-2}$, and assume
  the anisotropy profile given by eq.~(\ref{eq:betagamma}), with
  $\beta_{\infty}=0.36$ and $A=0.62$. A dotted line shows the best-fit
  linear relations, $\chi=2.1-1.16\, \alpha$.}
\label{FigAlphaChi} 
\end{figure}

Finally, our results imply a strong correlation between the
best-fitting Einasto's $\alpha$ and ``critical'' $\chi$ parameters for
any given halo. We show this in Fig.~\ref{FigAlphaChi}, which suggests
that the two parameters are essentially equivalent, and linked by a
simple linear relation, $\chi=2.1-1.16\, \alpha$. At the resolution of
current simulations, it does not seem possible to decide whether
Einasto or power-law PPSDs provide a better description of the
spherically-averaged structure of simulated CDM halos.

%%%%%%%%%%%%%%%%%%%%%%%%%%%%%%%%%%%%%%%%%%%%%%%%%%%%%%%%%%%%%%                                
%%%%%%%%%%%%%%%%%%%%%%%%%%%%%%%%%%%%%%%%%%%%%%%%%%%%%%%%%%%%%%          

\section{Summary and Conclusions}
\label{SecConc}

We use a sample of well-resolved equilibrium systems selected from the
MS-II simulation in order to study the density and pseudo-phase-space
density (PPSD) profiles of CDM halos. In particular, we explore the
relation between Einasto profiles (often used to parameterize
$\rho(r)$) and power-law PPSD profiles, which are often used to
construct equilibrium models of CDM halos.

We solve Jeans' equations to show that the PPSD profiles of Einasto
halos are close to power-laws, and that, conversely, the density
profiles of power-law PPSD models can be fitted by Einasto laws over a
wide radial range. The two descriptions differ, however, in their
inner asymptotic behaviour, although significant differences are
only expected at radii well inside $1\%$ of the scale radius,
$r_{-2}$, and are therefore beyond the reach of current simulations.

Our analysis of the MS-II halo sample shows that, over the resolved
radial range, Einasto and power-law PPSD models
provide an equally good description of the spherically-averaged
structure of simulated CDM halos, provided that the
Einasto parameter $\alpha$ and the power-law exponent $\chi$ are
allowed to float freely when fitting their radial structure. The strong 
correlation between best-fit $\alpha$ and $\chi$ parameters implies that they
constitute, in practice, equivalent measures of the shape of the mass
profile. These results confirm earlier suggestions that halo
structure is not strictly self-similar, and that a ``shape'' parameter
that varies from halo to halo is needed to characterize fully the
structure of CDM halos.

The spherically-averaged mass profiles of equilibrium CDM halos thus seem to
be fully specified by three parameters: two physical scalings,
$r_{-2}$ and $\rho_{-2}$, and a shape parameter, $\alpha$ or $\chi$. Where CDM
halos lie in this three-dimensional parameter space is likely linked
to each system's evolutionary history. Exploring the details
of these relations remains a pending issue, but one that can be
profitably addressed through large-scale numerical efforts such as the
Millennium Simulation series. We expect to report progress in this
area in the near future.

%%%%%%%%%%%%%%%%%%%%%%%%%%%%%%%%%%%%%%%%%%%%%%%%%%%%%%%%%%%%%%                                
%%%%%%%%%%%%%%%%%%%%%%%%%%%%%%%%%%%%%%%%%%%%%%%%%%%%%%%%%%%%%%          

\section*{acknowledgements}
The Millennium-II Simulation was carried out as part of the programme of the Virgo 
Consortium on the Regatta and VIP supercomputers at the Computing Centre of the
Max-Planck Society in Garching. JFN acknowledges support from the Canadian Institute 
for Advanced Research. CSF acknowledges a Royal Society Wolfson Research Merit Award. 
This work was supported in part by an STFC rolling grant to the ICC.

%%%%%%%%%%%%%%%%%%%%%%%%%%%%%%%%%%%%%%%%%%%%%%%%%%%                                                           
\begin{center}
\begin{table}
\caption{Parameters describing the velocity anisotropy profiles (eq.~(\ref{eq:betagamma}))}
\begin{tabular}{c c c}\hline \hline
$\langle\alpha\rangle$      & $\beta_{\infty}$  & $A$  \\ \hline
0.132     & 0.253  & 0.720  \\
0.178     & 0.313  & 0.656  \\
0.230     & 0.361  & 0.618  \\ \hline
\end{tabular}
\label{tab1}
\end{table}
\end{center}
%%%%%%%%%%%%%%%%%%%%%%%%%%%%%%%%%%%%%%%%%%%%%%

%\bibliographystyle{mn2e}
%\bibliography{paper}

\bsp
\label{lastpage}

\bibliographystyle{mn2e}
\bibliography{paper}

\end{document}